\documentclass[aps,letterpaper,
               final,twocolumn,
               superscriptaddress,
               showkeys,showpacs]
               {revtex4}
\usepackage{graphics}
\usepackage{graphicx,natbib,amssymb}
\usepackage[thinspace,squaren]{SIunits}
\def\full{\protect\mbox{------}}
\def\kesik{\protect\mbox{--\, --\, --}}
\def\dd{{\rm d}}

\def\dd{{\mathrm{d}}}

\def\full{\protect\mbox{------}}
\def\kesik{\protect\mbox{-\, -\, -\, -}}

\begin{document}
\title{Numerical Extraction of Distributions of Space-charge and Polarization from Laser Intensity Modulation Method}
\author{\firstname Enis \surname Tuncer}
\email{enis.tuncer@physics.org}
\affiliation{Applied Condensed-Matter Physics, University of Potsdam, D-14469 Potsdam, Germany}
\author{\firstname Sidney B. \surname Lang}
\affiliation{Department of Chemical Engineering, Ben-Gurion University of the Negev, 84105 Beer-Sheva, Israel}
\date{\today}

\begin{abstract}
The  Fredholm integral equation of the laser intensity modulation method is solved with the application of the Monte Carlo technique and a least-squares solver. The  numerical procedure is tested on simulated data.
\end{abstract}
\maketitle

Space-charge and polarization distributions in materials can be determined by the  Laser Intensity Modulation Method (LIMM), which was proposed by \citet{LangDasGupta} \citep[see Ref.][for a recent review on the LIMM]{LangIEEE2004}. In the LIMM experiment, a thin sample with electrodes on both surfaces is irradiated with a modulated laser. The light is absorbed at the irradiated electrode and the heat diffuses into the sample. Space-charge and polarization in the material respond to a non-uniform temperature distribution, and produce a pyroelectric current density $j$. The real $j_r$ and imaginary $j_i$ parts of the current are recorded for each modulation frequency $\nu$. The typical frequency range is between $10\ \hertz$ and $100\ \kilo\hertz$. \citet{LangIEEE2004} and \citet{Mellinger2004} have recently illustrated how to reconstruct the space-charge and polarization profiles with polynomial and Tikhonov\cite{Tikhonov} regularizations, respectively. In this letter, an alternative numerical method is presented and tested. 

The measured current density in  LIMM is expressed as a Fredholm integral equation\cite{LangIEEE2004},
\begin{eqnarray}
  \label{eq:1}
  j(\nu)=j_r+\imath j_i=\imath\ 2\pi\nu s^{-1}\int_0^s {\sf g}(z) \mathfrak{T}(\nu,z)\dd z
\end{eqnarray}
where ${\sf g}(z)$ is the unknown distribution of space charge or polarization, $\mathfrak{T}(\nu,z)$ is the solution of one-dimensional heat-conduction equation, $z$ is the coordinate in the thickness direction, and $\imath=\sqrt{-1}$. The integral in Eq.~(\ref{eq:1}) is evaluated over the samples thickness $s$. For a free standing film, $\mathfrak{T}(\nu,z)$ is\cite{Mellinger2004,Bloss1994} 
\begin{eqnarray}
  \label{eq:2}
  \mathfrak{T}(\nu,z)=A(\kappa\beta)^{-1}\cosh[\beta(s-z)] {\rm csch}(\beta s).
\end{eqnarray}
Here, $A$ is a constant, $\kappa$ is the thermal conductivity of the sample. $\beta$ is the complex thermal wave [$\beta=(1+\imath)\sqrt{\pi\nu D^{-1}}$ and $D$ is the thermal diffusivity of the material].

A numerical method based on the Monte Carlo method  have been proposed by \citet{Tuncer2000b} to solve integral equations in the form of Eq.~(\ref{eq:1}). The method has previously been applied to extract the distribution of relaxation times from dielectric spectroscopy data \cite{Tuncer2000b,Tuncer2004a} and the spectral density function (distribution) of dielectric mixtures \cite{Tuncer2D}. In those problems, the distributions were probability densities of relaxation times and spectral parameters. Therefore, a constrained least-squares algorithm was implemented which yielded only positive valued outputs. In the present problem, 
the space-charge and the polarization distributions ${\sf g}$ can have both positive and negative magnitudes. As a result, we employ a similar numerical procedure as before \cite{Tuncer2000b}. However, here a linear least-squares solver is adopted without any {\em a-priori} assumptions instead of the constrained least-squares method. 
\begin{figure}[t]
  \centering
  \includegraphics[width=8.5cm]{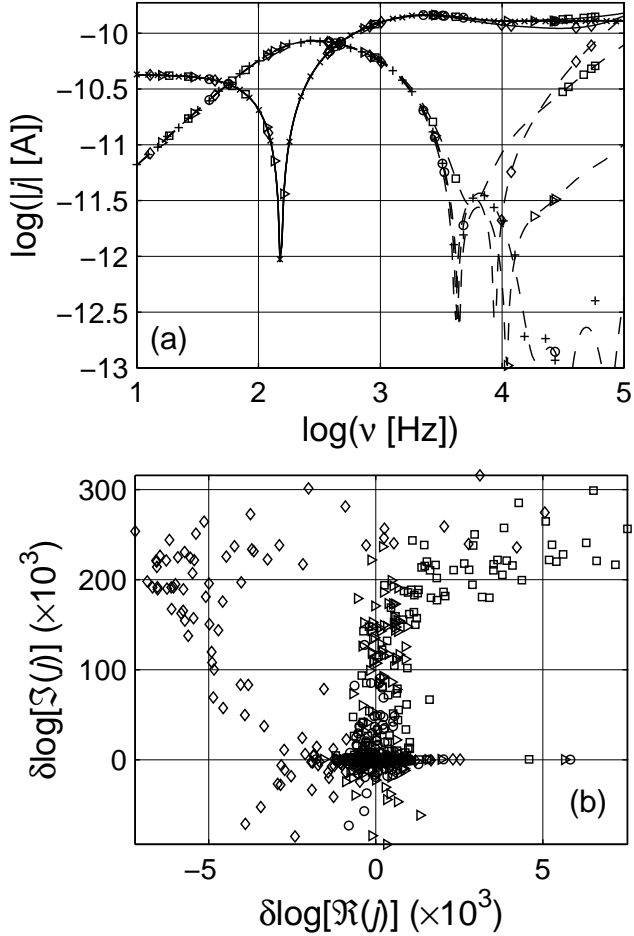}
  \caption{(a) Simulated current and the current reconstructed with the least-squares. The symbols $(\times)$ and $(+)$ represent the real (\full) and imaginary (\kesik) parts of the simulated current density $j$. The open symbols are the reconstructed currents with different number of bins; ($\circ$) $M/2$, ($\triangleright$) $M/4$, ($\Box$) $M/8$ and ($\diamond$) $M/16$, where $M$ is the number of data points. (b) Relative error in the logarithm of currents calculated. The symbols correspond to the reconstructed currents in (a).  The reconstructed data are calculated with $M/n=1.1$ and $n\times N\simeq10^{4}$.\label{fig:current}}  
\end{figure}

The numerical procedure proposed is as follows,
\begin{enumerate}
\item Eq.~(\ref{eq:1}) is written as a summation over some number $n$ of $z$ values for each experimental frequency point $\nu_m$ ($m=1,\dots,M$, $M$ is the number of experimental data points),
  \begin{eqnarray}
    \label{eq:3}
    j(\nu_m)=\imath\ 2\pi\nu_m s^{-1}\sum_{n<M} {\sf g}(z_n) \mathfrak{T}(\nu_m,z_n).
  \end{eqnarray}  
\item The $z_n$ values are preselected randomly from a linear distribution, $0<z_n<s$.
\item Eq.~(\ref{eq:3}) is written as a matrix equation,
  \begin{eqnarray}
    \label{eq:4}
    \mathbf{T} \mathbf{g} = \mathbf{j},
  \end{eqnarray}
and solved for $\mathbf{g}$. In Eq.~(\ref{eq:4}), $\mathbf{T}$ is the $m\times n$ {\em `kernel'} matrix, where $\mathbf{T}=\mathfrak{T}(\nu_m,z_n)$, $\mathbf{g}$ is a column vector with length $n$ ($\mathbf{g}={\sf g}_n$), and $\mathbf{j}$ is a column vector with length $m$. 
\item The minimization then yields the desired distribution $\mathbf{g}$
  \begin{eqnarray}
    \label{eq:5}
    R^2=\sum_m [j(\nu_m)-\mathbf{Tg}]^2.
  \end{eqnarray}
{\em Since the problem is ill-conditioned and has many solutions, one single minimization does not lead to a reasonable (trustworthy) solution.} However, we  eliminate this drawback by running the minimization for many times with newly selected $z$-values. The resulting sets of ${\sf g}$ values are finally averaged. 
The steps 1 through 4 are executed $N$ times with different sets of $z$ (as in a Monte Carlo technique). The $z_n$ and ${\sf g}_n$ values are recorded at each step, and the final distribution is obtained from the ${\sf g}(z)$ histogram. The histogram is obtained by dividing the $z$-axis into a number of bins (channels), and the average of ${\sf g}$ values is calculated in each channel. It should be noted that the introduction of the Monte Carlo technique helps us to select independent sets of $z_n$, and the $z$-axis is continuous in this approach, which is not the case in most regularization methods. The channel widths give the resolution of the LIMM experiment.
\end{enumerate}
\begin{figure}[t]
  \centering
  \includegraphics[width=8.5cm]{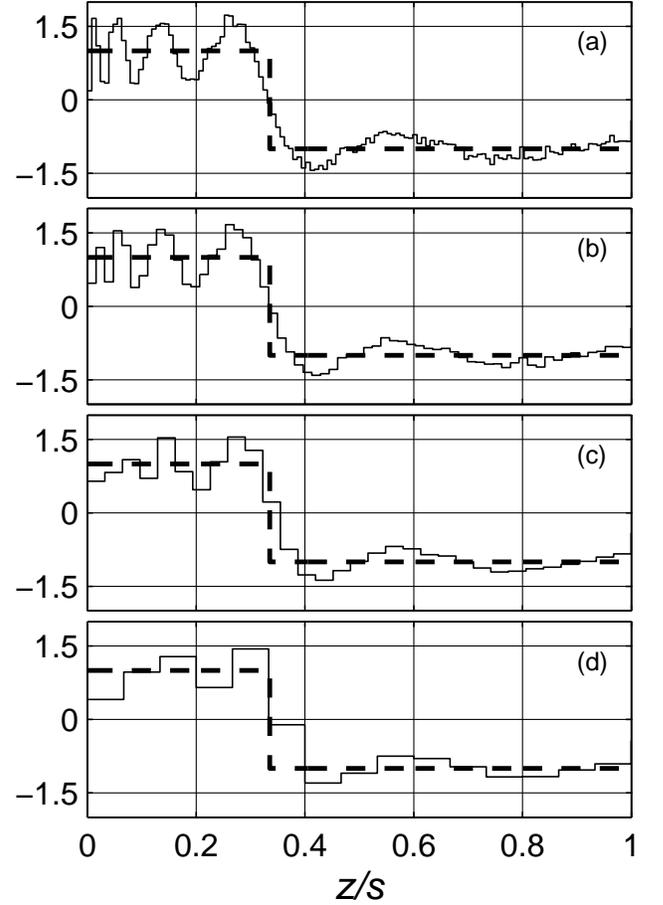}
  \caption{Normalized polarizations ${\sf g(z)}/\max[{\sf g}_0(z)]$ calculated with the least-squares for four different numbers of bins: (a) $M/2$; (b) $M/4$; (c) $M/8$; (d) $M/16$, where $M$ is the number of data points. The thick dashed line (\kesik) represents the original polarization distribution. The distributions are calculated with $M/n=1.1$ and $n\times N\simeq10^{4}$.\label{fig:polarization}}
\end{figure}

The proposed method is sensitive to the number $n$ of initial unknowns and the number $N$ of the Monte Carlo cycles. Best results are obtained when $M/n<1.2$ and $n\times N>5\times10^3$. It is observed that for $N>512$, the results do not improve significantly, and the computation time increases enormously. The reconstructed curves below are calculated with $M/n=1.1$ and $n\times N\simeq10^{4}$, which took less than $2\ \minute$ on a Pentium 4 ($2.4\ \giga\hertz$ {\sf LinuxPc}) with $1\giga \textrm{byte}$ memory. 

The original data is generated by Eq.~(\ref{eq:2}) by considering a step polarization distribution ${\sf g}_0(z)$ with $\Delta z=4\times10^{-8}$,
\begin{eqnarray}
  \label{eq:6}
  {\sf g}_0(z)=\left\{ \begin{array}{ll}
      +1\times{\Delta z} &\quad  z\le s/3\\
      -1\times{\Delta z} &\quad  z>s/3
    \end{array} \right.
\end{eqnarray}
We assume a $20\ \micro\meter$ thick sample with thermal parameters $\kappa=0.948\ \watt\reciprocal{(\meter\usk\kelvin)}$ and $D=6.09\times 10^{-8}\ \squaremetrepersecondnp$, and $A=1$ in Eq.~(\ref{eq:2}). The original data has $256$ frequency points in logarithmic scale, $M=256$. In the calculations, both the real and the imaginary parts of the current are used. We add both $1\%$ Gaussian noise and a white noise of level of $0.1\ \pico\ampere$ to the real and the imaginary parts of the generated data.

In Fig.~\ref{fig:current}, the simulated and the reconstructed currents are presented together with the relative error in the logarithm of the currents. In Fig.~\ref{fig:current}a, the real and the imaginary parts of the current density $j$ are denoted by $(x)$ and $(+)$ symbols, respectively.  The proposed method is very successful in restoring the original data. The data is reconstructed with various bin sizes. Since the imaginary part of the data has lower values than the real part, the relative error is larger than in the real part, and the restored data at high frequencies differ from the original data in the log-log presentation, as shown in Fig.~\ref{fig:current}b. Best agreement between generated and the reconstructed data is obtained for the largest number of bins considered, $M/2$ corresponding to the smallest channel width $2s/M$.

The normalized original and calculated distributions are presented in Fig.~\ref{fig:polarization}. In the calculations different bin widths are used in order to average the results. The $z$-axis is divided into $2^{-i}M$ divisions with $i=1,\dots,4$. The numerical method is able to resolve the step; however, there are oscillations when $z<s/3$ which are not present in the original distribution. This is due to the applied least-squares solver, which yielded large positive and negative ${\sf g}$ values when two randomly selected $z$  were very close. This effect could be removed by using smaller number of bins as presented in Fig.~\ref{fig:polarization}.
An alternative way to reduce the oscilations is by implementing a constrained least-squares algorithm in which the results from the numerical analysis with the least-squares is used as an {\em a-priori} assumption. 
In such an approach, the signs of the calculated ${\sf g}$ in Fig.~\ref{fig:polarization} are used as a pre-distribution for the sign of the polarization distribution in a numerical method based on the constrained least-squares algorithm~\cite{Tuncer2000b,Tuncer2004a,Tuncer2D}. 
The averaging of ${\sf g}$ values is performed by considering various bin sizes. The obtained distributions become closer to the original one as we increase the channel width (decrease the number of bins). Note that the resolution in  LIMM is material dependent and for this particular generated data set, it is around $0.5\ \micro\meter$, corresponding to approximately $M/8$ number of bins, as in Figs.~\ref{fig:polarization}c. As a results, it is not meaningful or necessary to use smaller channel widths to improve the polarization distributions. 

In conclusion, we have solved the inverse problem of the LIMM equation using the Monte Carlo and linear least-squares methods simultaneously. This method is  a valuable alternative to those used in the literature. 
~\\

 We would like to express our thanks to Dr Axel Mellinger for the discussions and his suggestion to generate the data.


\end{document}